# Probing Inhomogeneous Cuprate Superconductivity by Terahertz Josephson Echo Spectroscopy

A. Liu[1,2], D. Pavicevic[1], M. H. Michael[1], A. G. Salvador[3], P. E. Dolgirev[4], M. Fechner[1], A. S. Disa[1], P. M. Lozano[2,5], Q. Li[2,5], G. D. Gu[2], E. Demler[3], A. Cavalleri[1,6]

[1]Max Planck Institute for the Structure and Dynamics of Matter, Hamburg, Germany
[2]Condensed Matter Physics and Materials Science Division, Brookhaven National Laboratory, New York, USA
[3]Institute for Theoretical Physics, ETH Zurich, Zurich, Switzerland
[4]Department of Physics, Harvard University, Massachusetts, USA
[5]Department of Physics and Astronomy, Stony Brook University, New York, USA
[6]Department of Physics, University of Oxford, Oxford, United Kingdom

**Inhomogeneities play a crucial role in determining the properties of quantum materials. Yet methods that can measure these inhomogeneities are few, and apply to only a fraction of the relevant microscopic phenomena. For example, the electronic properties of cuprate materials are known to be inhomogeneous over nanometer length scales, although questions remain about how such disorder influences supercurrents and their dynamics. Here, two-dimensional terahertz spectroscopy is used to study interlayer superconducting tunneling in near-optimally-doped $La_{1.83}Sr_{0.17}CuO_4$. We isolate a 2 THz Josephson echo signal with which we disentangle intrinsic lifetime broadening from extrinsic inhomogeneous broadening. We find that the Josephson plasmons are only weakly inhomogeneously broadened, with an inhomogeneous linewidth that is three times smaller than their intrinsic lifetime broadening. This extrinsic broadening remains constant up to $0.7T_c$, above which it is overcome by the thermally-increased lifetime broadening. Crucially, the effects of disorder on the Josephson plasma resonance are nearly two orders of magnitude smaller than the in-plane variations in the superconducting gap in this compound, which have been previously documented using Scanning Tunnelling Microscopy (STM) measurements. Hence, even in the presence of significant disorder in the superfluid density, the finite frequency interlayer charge fluctuations exhibit dramatically reduced inhomogeneous broadening. We present a model that relates disorder in the superfluid density to the observed lifetimes.**

Unconventional superconductivity in cuprates emerges when either electrons or holes are doped into the insulating parent compound. Because superconductivity is optimized far away from optimal stoichiometric composition, disorder of the host lattice is unavoidable and may result in significant electronic inhomogeneities [1, 2]. Measurements of the electronic properties of cuprates using scanning tunneling microscopy [3] reveal disorder of the superconducting gap on nanometer length scales [4, 5], whose variations are further correlated with the distribution of dopant atoms [6]. Because coherence lengths in cuprates are comparable to the length scales of dopant inhomogeneities [7], the effect of lattice disorder on the Cooper pairs is not obvious. Additional probes of the role of these inhomogeneities on the superconducting order parameter are therefore necessary to reach a complete picture of the role of disorder in these materials.

The tunneling resonance arising from Josephson coupling between adjacent superconducting $CuO_2$ planes, the so-called Josephson plasma resonance [8, 9], provides a direct measure of the superconducting order parameter and of the c-axis transport [10, 11]. The Josephson plasma resonance has been extensively studied by linear spectroscopy, in which the c-axis reflectivity exhibits a pronounced plasma edge in the superconducting state (shown in Figure 1a). van der Marel and Tsvetkov first pointed out [12] that a distribution of Josephson plasma frequencies, reflecting disorder of the superconducting order parameter, manifests as a distortion of the Josephson plasmon loss function [13]. Corresponding distortion of the reflectivity edge profile was then exploited by Dordevic et al. [14] to quantify the underlying disorder in $La_{2-x}Sr_xCuO_4$.

However, the above approach suffers from various shortcomings. For example, in [14] the lineshape fits based on a two-fluid model are generally not completely constrained and required assumptions on both the normal fluid conductivity and the functional form of the plasma frequency distribution. Reliable fits are also only possible at low temperatures, in which a sharp plasma edge is observed.

To circumvent these limitations, techniques capable of resolving higher-order plasmon correlation functions are required. These higher-order plasmon correlations may be measured by nonlinear spectroscopy [15, 16], one example of which is shown schematically in Figure 1b where two excitation fields $E_A$ and $E_B$ cooperatively generate a nonlinear electric field $E_{NL}$. The depicted nonlinear process is governed by a third-order optical response $\chi_\varphi^{(3)}$, with multiple frequency and wavevector components. These may be resolved in a two-dimensional (2-D) spectrum [17, 18] (shown in Figure 1c) by Fourier transforming $E_{NL}$ along the inter-pulse time-delay $\tau$ and the emission time t. So-called 2-D spectroscopy has been implemented at terahertz frequencies in collinear excitation geometries to study a variety of material systems [19, 20, 21, 22, 23, 24, 25,

26, 27, 28] [29], including superconductors more recently [30, 31, 32]. In the non-collinear excitation geometry shown here, different components of $\chi_\varphi^{(3)}$, corresponding to different peaks in the 2-D spectrum, are emitted in unique phase-matched directions.

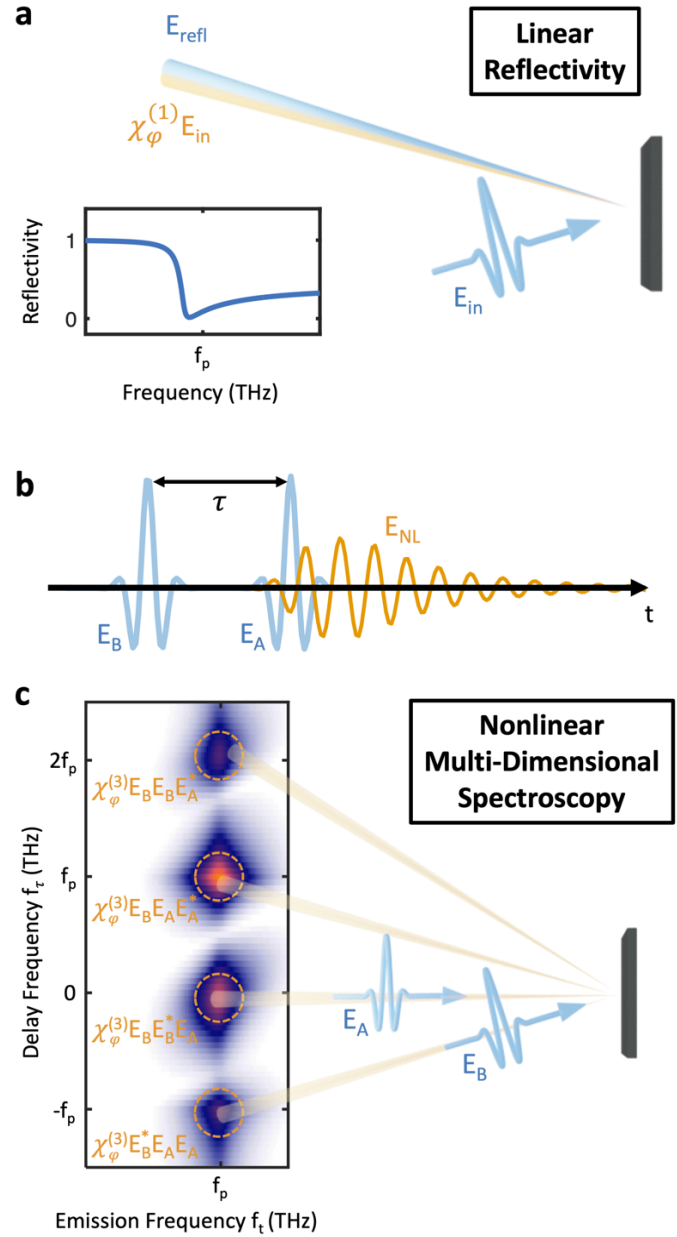

**Fig. 1. Linear and nonlinear spectroscopies of plasma resonances. a.** *Schematic of a linear reflectivity measurement, in which an incident field $E_{in}$ induces a linear response $\chi_\varphi^{(1)}$. The co-propagating reflected field $E_{refl}$ and the linear response $\chi_\varphi^{(1)}$ interfere to produce a characteristic edge in the reflectivity shown inset.* **b.** *Generation of a nonlinear electric field $E_{NL}$ by two excitation pulses $E_A$ and $E_B$ of comparable amplitude. The amplitude and phase of $E_{NL}$ depend on the inter-pulse time-delay $\tau$.* **c.** *Schematic of nonlinear multi-dimensional spectroscopy, in which Fourier transform of $E_{NL}$ along the two time variables $\{\tau,t\}$ resolves a nonlinear optical response $\chi_\varphi^{(3)}$ in a two-dimensional spectrum. In the non-collinear excitation geometry shown, various components of $\chi_\varphi^{(3)}$ are emitted in unique phase-matched directions shown.*

The non-collinear phase-matching geometry is shown in Figure 2a, in which two quantities are conserved. In addition to the momentum magnitude determined by the Josephson plasma frequency, only in-plane momentum is conserved due to the interface [33]. For fixed excitation and detection geometry, the nonlinearity arriving at the detector may be chosen by rotating the sample and thereby the in-plane momentum of each excitation beam (shown in Figure 2b).

We demonstrate this principle in Figure 2c on near-optimally-doped $La_{2-x}Sr_xCuO_4$ (x = 0.17, LSCO), which exhibits a Josephson plasma resonance with resonance frequency $f_p \approx 2$ THz at temperatures far below the phase transition (T $\ll$ $T_c$). All four components of $\chi_\varphi^{(3)}$ that radiate at the Josephson plasma frequency [34, 35] are resolved at their respective phase-matching conditions. Here, we focus on the nonlinearity appearing at $(f_t, f_\tau)$ = (2,-2) THz, which corresponds to a terahertz frequency 'Josephson echo'.

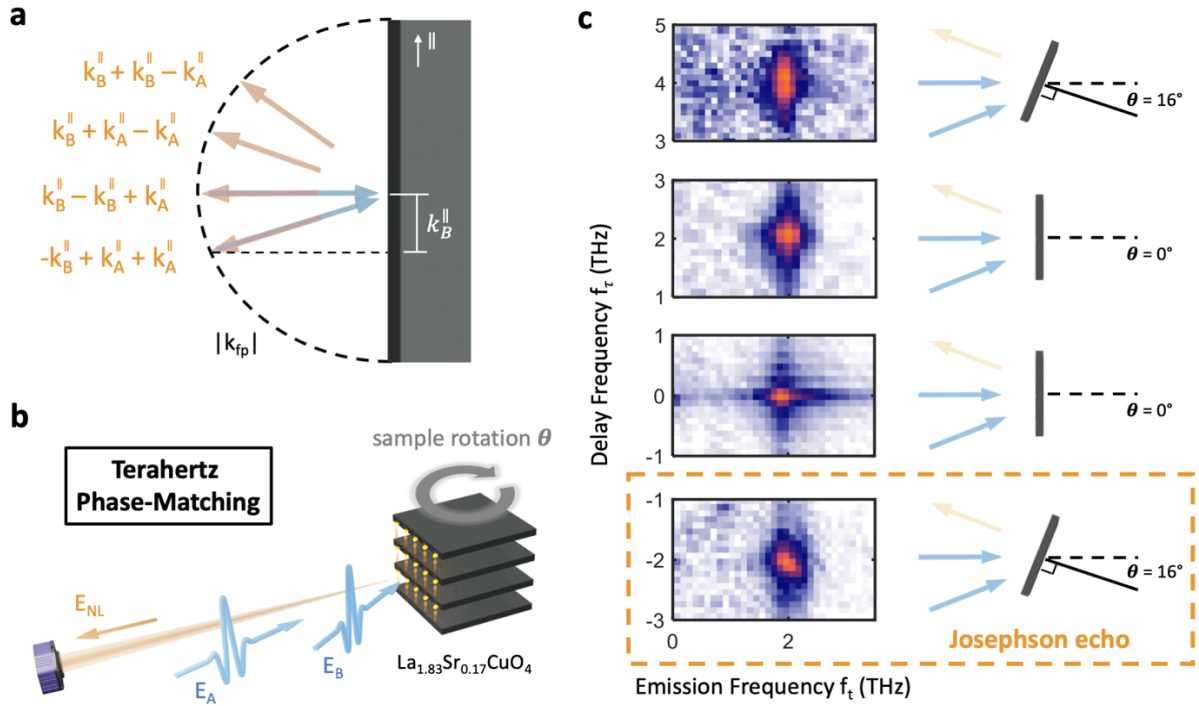

*Fig. 2. Phase-matching resolves individual plasmon nonlinearities. **a.** Schematic of wave-vector phase-matching, in which momentum is only conserved along the in-plane (∥) direction and in its magnitude $|k_{fp}|$. **b.** Sample rotation changes the in-plane momenta of each excitation field and determines the nonlinearity that arrives at the detector. **c.** Experimental measurements of the constituent components of $\chi_\varphi^{(3)}$ in $La_{1.83}Sr_{0.17}CuO_4$ are shown with the corresponding sample rotation angles. The $(f_t, f_\tau)$ = (2,±2) THz peaks are measured with $E_B$ arriving first, followed by $E_A$. The (2,0) and (2,4) THz peaks are measured with $E_A$ arriving first, followed by $E_B$. The (2,-2) THz peak corresponds to an emitted Josephson echo.*

The advantage of Josephson echoes in measuring disorder is illustrated by Figure 3, in which two cartoons of homogeneous and disordered interlayer tunneling are shown in Figure 3a. If intrinsic spectral broadening of the resonance (for example due to quasiparticle screening) is comparable to a distribution of Josephson plasma frequencies (due to variations in the interlayer tunneling response), comparable optical responses are observed by linear optical spectroscopy (Figure 3b) and one cannot distinguish between these two physically distinct situations.

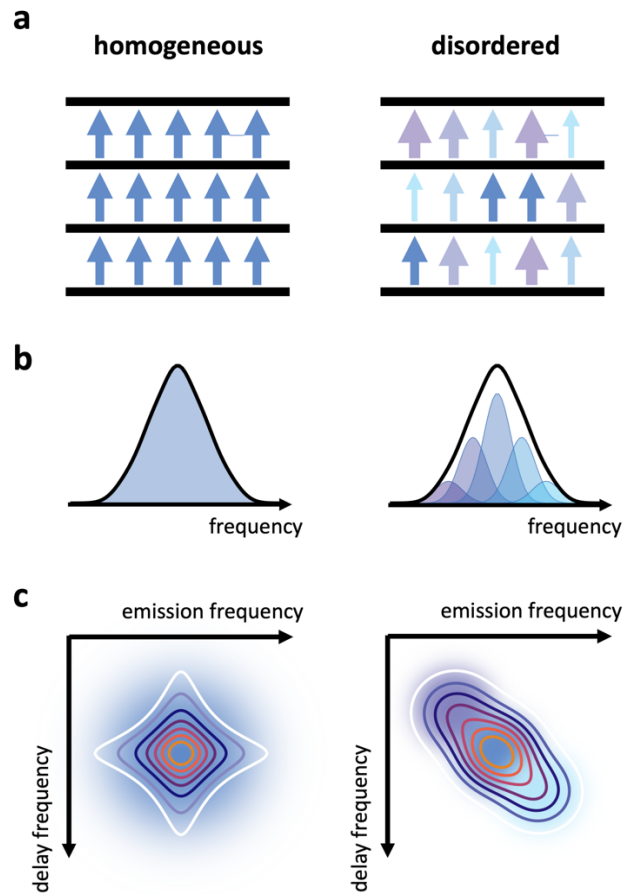

*Fig. 3. Disorder in schematic one- and two-dimensional spectra. a. Cartoon of homogeneous and disordered Josephson tunneling in a layered superconductor. b. Homogeneous and disordered Josephson plasma resonance in one-dimensional spectra, which exhibit ambiguous lineshapes (black curves) in the presence of comparable levels of intrinsic and disorder broadening. c. Homogeneous and disordered Josephson plasma resonance in two-dimensional Josephson echo spectra, in which disorder is evident through asymmetry of the Josephson echo peak.*

This ambiguity is eliminated in a two-dimensional spectrum, specifically in the spectral lineshape of the Josephson echo peak [36]. In the case of a homogeneous Josephson plasma resonance, the peak is symmetric with identical peak widths along the two frequency axes. In the presence of disorder however, the Josephson echo peak develops a marked asymmetry from projecting disorder line-broadening into an orthogonal direction from the intrinsic linewidth of the resonance.

Measured 2-D spectra of the Josephson echo in LSCO are shown in Figure 4 for increasing temperatures approaching $T_c \approx 36$ K. At the lowest measured temperature of 6 K, the Josephson echo peak is asymmetric with a characteristic 'almond' shape that reveals a finite degree of disorder [36, 37]. At higher temperatures the peak becomes increasingly symmetric, suggesting a crossover into the regime of dominant intrinsic line-broadening. Next, we quantify this statement by extracting the individual intrinsic and disorder contributions to the Josephson plasma resonance peak width.

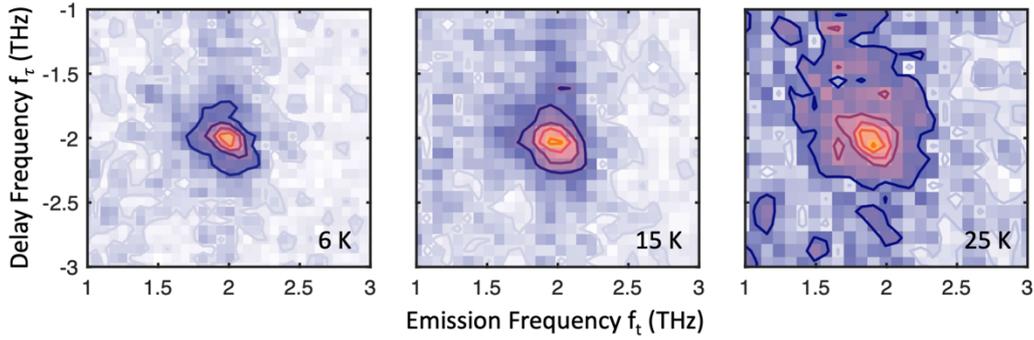

*Fig. 4. Two-dimensional Josephson echo spectra of La$_{1.83}$Sr$_{0.17}$CuO$_4$ for increasing temperatures. Two-dimensional spectra in the $(f_t, f_\tau) = (f_p, -f_p)$ quadrant with equal-value contours plotted on top of the raw data. At the lowest temperature of 6 K, 'almond' asymmetry of the peak is clearly observed that indicates the presence of disorder. At higher temperatures, the peak lineshape becomes increasingly symmetric due to a lesser relative importance of disorder.*

In 2-D spectra, the disorder linewidth $\gamma_{disorder}$ is projected into a characteristic direction along the 'diagonal' ($|f_t| = |f_\tau|$) line, in contrast to the intrinsic linewidth $\gamma_{intrinsic}$ which results in symmetric broadening. In the limit of dominant disorder ($\gamma_{disorder} \gg \gamma_{intrinsic}$), the lineshapes along the 'diagonal' and perpendicular 'anti-diagonal' directions are decoupled [36, 38] and depend solely on $\gamma_{disorder}$ and $\gamma_{intrinsic}$ respectively [36]. In the present case of comparable intrinsic and disorder broadening however, the two lineshapes must be simultaneously fit [37] to extract $\gamma_{disorder}$ and $\gamma_{intrinsic}$.

Slices of the Josephson echo peak at 6 K are plotted in Figure 5a, along the directions shown in the insets. The lineshape along the 'diagonal' direction (orange) is broader than the lineshape along the 'anti-diagonal' direction (blue), and the difference between the two peak widths indicates disorder line-broadening. Simultaneous fits of the two lineshapes (to functional forms described in the Supplemental Information) were performed, from which we extract values of $\gamma_{disorder} = 0.08$ THz and $\gamma_{intrinsic} = 0.38$ THz. This value for $\gamma_{disorder}$ is comparable to that reported by Dordevic et al. [14] from linear reflectivity, which was extracted under an assumed value for $\gamma_{intrinsic}$.

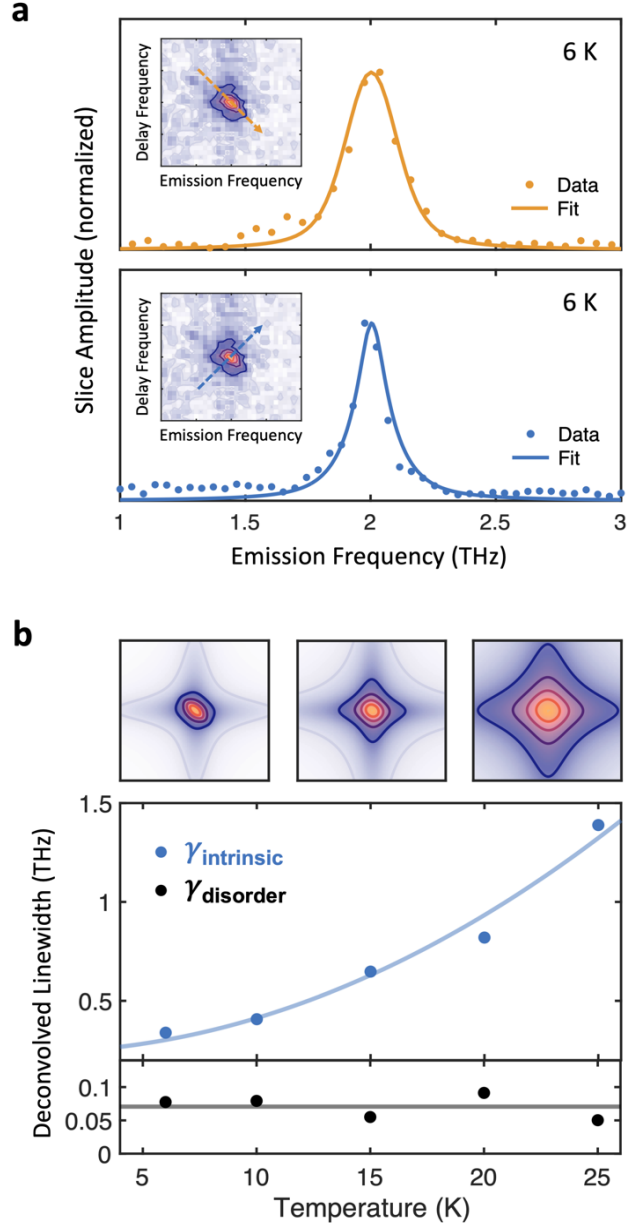

*Fig. 5. Fitting the Josephson echo peak lineshape to extract intrinsic and disorder linewidths. **a.** Slices of the Josephson echo peak at 6 K along the 'diagonal' (orange) and 'anti-diagonal' (blue) directions shown inset. The lineshapes of both slices are broadened by intrinsic damping and disorder, and fits of the two lineshapes are thus performed simultaneously to extract $\gamma_{disorder}$ and $\gamma_{intrinsic}$. **b.** Temperature dependence of $\gamma_{disorder}$ and $\gamma_{intrinsic}$, in which disorder of the Josephson plasma resonance remains roughly constant while its intrinsic damping increases rapidly with temperature. Lines are guides to the eye, and 2-D spectra shown above are simulated with the fitted parameters at 6 K, 15 K, and 25 K from left to right respectively.*

In the linear optical response, the reflectivity edge associated with superconductivity rapidly fades into a featureless background with increasing temperature [10]. However, 2-D terahertz spectroscopy distills the superfluid response from this normal fluid background and uniquely enables us to follow disorder in the superconducting transport even at temperatures approaching the phase transition. Extracted values of $\gamma_{disorder}$ and $\gamma_{intrinsic}$ are shown in Figure 5b as a function of increasing temperature. First, the intrinsic linewidth $\gamma_{intrinsic}$ increases rapidly with temperature due to thermally-activated quasiparticle excitations or perhaps even topological defects [39]. More surprising is the behavior of the disorder linewidth $\gamma_{\text{disorder}}$, which remains roughly constant up to a temperature of 25 K ≈ 0.7$T_c$. Above this temperature, the Josephson plasma resonance becomes lifetime-limited and the disorder becomes unmeasurable.

One may expect the disorder to increase with temperature as the coherence length (reflecting the spatial extent of the Cooper wavefunction [40]) shrinks [41]. The independence of $\gamma_{\text{disorder}}$ with respect to temperature thus raises the question of whether the disorder linewidth already reflects the full extent of the underlying electronic disorder. We therefore compare in Figure 6 the measured distribution of plasma frequencies (of standard deviation $\gamma_{\text{disorder}}$) to the previously measured superconducting gap distribution in LSCO [5], which is nearly two orders of magnitude broader. To relate these two scales of disorder, we assume that, due to the short coherence length [42] in LSCO [7], the superfluid density directly inherits identical disorder from the underlying gap fluctuations. In this scenario, disorder in the superconducting gap is related to $\gamma_{\text{disorder}}$ by the following relation (see Supplementary Information):

$$\frac{\sigma_\Delta^2}{\Delta_0^2} = \frac{\sigma_{|\psi|^2}^2}{(|\psi|^2)_0^2} = \frac{\sqrt{2\ln(2)}\,\gamma_{disorder}}{f_p} \frac{4 f_{ab}^2 d^2}{\pi^2 f_p^2 \xi^2}$$

where $d$ is the interlayer spacing, $f_{ab}$ is the in-plane plasma frequency, and $\xi$ is the correlation length of the disorder. Using the gap disorder parameters from [5], namely a mean superconducting gap $\Delta_0$ = 10 meV with standard deviation $\sigma_\Delta$ = 1.7 meV and an estimate for the correlation length $\xi$ ≈ 10 nm, we obtain $\gamma_{\text{disorder}}$ ≈ 80 GHz (see Supplementary Information), in excellent agreement with the measured values shown in Figure 5b. We thus conclude that spatial fluctuations in the superconducting gap directly induce commensurate fluctuations in the superfluid density that are then averaged over in the c-axis optical response.

We note that nonlinear multidimensional responses are still observed immediately above $T_c$, and fitting of their corresponding Josephson echo signal is currently prevented by signal amplitudes and signal-to-noise ratio. Suitable technical improvements will enable nonlinear probing of the partially coherent normal state, and may provide precious new information on the nature of the pseudogap phase. We also note how new and frequency-agile terahertz sources, which are becoming available due to improvements in laser instrumentation [43], will enable systematic measurements throughout the phase diagram of a single compound and across many families of cuprates, where frequencies of the plasma resonance vary between 100 GHz and 15 THz [8].

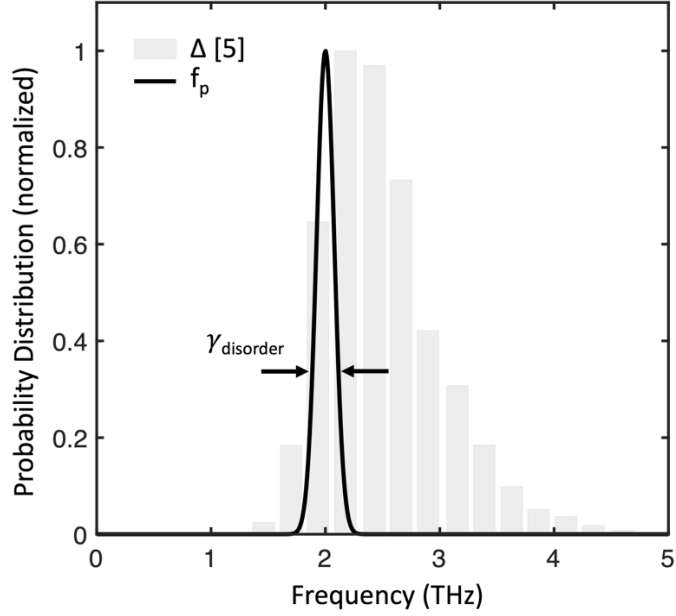

*Fig. 6. Comparison of disorder in the superconducting gap and in the superconducting transport. A probability distribution for the Josephson plasma frequencies $f_p$ (solid line) is compared to that of the superconducting gap Δ (shaded bars) measured using scanning tunneling microscopy [5], measured at 6 K and 4.2 K respectively. The standard deviation of the plasma frequency distribution $\gamma_{disorder}$ = 78 GHz is far smaller than the standard deviation of the superconducting gap $\gamma_\Delta$ = 2.8 THz.*

The present results provide also a useful working hypothesis to rationalize the enigmatic phenomenon of light-induced superconductivity in cuprates [44]. Two hallmarks of a superconducting c-axis response, a reflectivity edge and a 1/ω divergence in the inductive (imaginary) part of the optical conductivity, are observed transiently following resonant phonon excitation [45, 46]. Yet these indications of activated superconducting tunnelling in the out-of-plane responses are frequently accompanied by robust in-plane Ohmic responses, which likely reflects inhomogeneities in the photo-induced state. The results we present demonstrate robustness of the coherent c-axis optical response to disorder [39]. This observation lends credence to the scenario of an inhomogeneous non-equilibrium state that features local superconducting correlations even in the absence of long-range order.

In summary, we have made use of two-dimensional terahertz spectroscopy to quantify the role of disorder in the interlayer tunneling, and thereby in the superconducting condensate. By measuring the terahertz Josephson echo, we observe an interlayer tunneling response that is largely immune to the underlying electronic disorder, which remains true even as temperature approaches the phase transition. This demonstration of terahertz echoes [23] from a collective excitation provides us with a method to study inhomogeneities in a vast range of quantum materials, ranging from incipient ferroelectrics [47] to spin liquids [48]. Furthermore, the inherent ultrafast nature of this method makes it applicable to multi-dimensional probes of light-induced non-equilibrium phenomena [49, 50], to understand the role of disorder in transient states and their formation mechanisms.

# References

[1] P. M. Singer, A. W. Hunt and T. Imai, "63Cu NQR Evidence for Spatial Variation of Hole Concentration in La2-xSrxCuO4," *Phys. Rev. Lett.,* vol. 88, no. 4, 2002.

[2] E. S. Bozin, G. H. Kwei, H. Takagi and S. J. L. Billinge, "Neutron Diffraction Evidence of Microscopic Charge Inhomogeneities in the CuO2 Plane of Superconducting La2-xSrxCuO4 (0 ≤ x ≤ 0.30)," *Phys. Rev. Lett.,* vol. 84, no. 25, 2000.

[3] A. Yazdani, E. H. da Silva Neto and P. Aynajian, "Spectroscopic imaging of strongly correlated electronic states," *Annu. Rev. Condens. Matter Phys.,* vol. 7, no. 1, pp. 11-13, 2016.

[4] S. H. Pan, J. P. O'Neal, R. L. Badzey, C. Chamon, H. Ding, J. R. Engelbrecht, Z. Wang, H. Eisaki, S. Uchida, A. K. Gupta, K.-W. Ng, E. W. Hudson, K. M. Lang and J. C. Davis, "Microscopic electronic inhomogeneity in the high-Tc superconductor Bi2Sr2CaCu2O8+x," *Nature,* vol. 413, pp. 282-285, 2001.

[5] T. Kato, S. Okitsu and H. Sakata, "Inhomogeneous electronic states of La2-xSrxCuO4 probed by scanning tunneling spectroscopy," *Phys. Rev. B,* vol. 72, no. 14, 2005.

[6] K. McElroy, J. Lee, J. A. Slezak, D.-H. Lee, H. Eisaki, S. Uchida and J. C. Davis, "Atomic-scale sources and mechanism of nanoscale electronic disorder in Bi2Sr2CaCu2O8+δ," *Science,* vol. 309, no. 5737, pp. 1048-1052, 2005.

[7] R. Wesche, Physical properties of high-temperature superconductors, Hoboken: Wiley-VCH, 2015.

[8] Y. Laplace and A. Cavalleri, "Josephson plasmonics in layered superconductors," *Adv. Phys. X,* vol. 1, no. 3, pp. 387-411, 2016.

[9] F. Gabriele, C. Castellani and L. Benfatto, "Generalized plasma waves in layered superconductors: a unified approach," *Phys. Rev. Res.,* vol. 4, no. 2, 2022.

[10] K. Tamasaku, Y. Nakamura and S. Uchida, "Charge dynamics across the CuO2 planes in La2-xSrxCuO4," *Phys. Rev. Lett.,* vol. 69, no. 9, 1992.

[11] D. N. Basov and T. Timusk, "Electrodynamics of high-Tc superconductors," *Rev. Mod. Phys.,* vol. 77, no. 2, 2005.

[12] D. van der Marel and A. Tsvetkov, "Transverse optical plasmons in layered superconductors," *Czech. J. Phys.,* vol. 46, pp. 3165-3168, 1996.

[13] M. Dressel and G. Gruner, Electrodynamics of Solids, Cambridge: Cambridge University Press, 2010.

[14] S. V. Dordevic, S. Komiya, Y. Ando and D. N. Basov, *Phys. Rev. Lett.,* vol. 91, no. 16, 2003.

[15] S. Mukamel, Principles of Nonlinear Optical Spectroscopy, Oxford: Oxford University Press, 1995.

[16] D. B. Turner and K. A. Nelson, "Coherent measurements of high-order electronic correlations in quantum wells," *Nature,* vol. 466, pp. 1089-1092, 2010.

[17] S. Mukamel, Y. Tanimura and P. Hamm, "Coherent multidimensional optical spectroscopy," *Acc. Chem. Res.,* vol. 42, no. 9, pp. 1207-1209, 2009.

[18] K. Reimann, M. Woerner and T. Elsaesser, "Two-dimensional terahertz spectroscopy of condensed-phase molecular systems," *J. Chem. Phys.,* vol. 154, no. 12, 2021.

[19] J. Savolainen, S. Ahmed and P. Hamm, "Two-dimensional Raman-terahertz spectroscopy of water," *PNAS,* vol. 110, no. 51, 2013.

[20] T. Maag, A. Bayer, S. Baierl, M. Hohenleutner, T. Korn, C. Schuller, D. Schuh, D. Bougeard, C. Lange, R. Huber, M. Mootz, J. E. Sipe, S. W. Koch and M. Kira, "Coherent cyclotron motion beyond Kohn's theorem," *Nat. Phys.,* vol. 12, pp. 119-123, 2016.

[21] C. Somma, G. Folpini, K. Reimann, M. Woerner and T. Elsaesser, "Two-phonon quantum coherences in indium antimonide studied by nonlinear two-dimensional terahertz spectroscopy," *Phys. Rev. Lett.,* vol. 116, no. 17, 2016.

[22] M. Grechko, T. Hasegawa, F. D'Angelo, H. Ito, D. Turchinovich, Y. Nagata and M. Bonn, "Coupling between intra- and intermolecular motions in liquid water revealed by two-dimensional terahertz-infrared-visible spectroscopy," *Nat. Commun.,* vol. 885, no. 2018, p. 9.

[23] J. Lu, Y. Zhang, H. Y. Hwang, B. K. Ofori-Okai, S. Fleischer and K. A. Nelson, "Nonlinear two-dimensional terahertz photon echo and rotational spectroscopy in the gas phase," *PNAS,* vol. 113, no. 42, 2016.

[24] C. L. Johnson, B. E. Knighton and J. A. Johnson, "Distinguishing nonlinear terahertz excitation pathways with two-dimensional spectroscopy," *Phys. Rev. Lett.,* vol. 122, no. 7, 2019.

[25] F. Mahmood, D. Chaudhuri, S. Gopalkrishnan, R. Nandkishore and N. P. Armitage, "Observation of a marginal Fermi glass," *Nat. Phys.,* vol. 17, pp. 627-631, 2021.

[26] S. Pal, N. Strkalj, C. J. Yang, M. C. Weber, M. Trassin, M. Woerner and M. Fiebig, "Origin of terahertz soft-mode nonlinearities in ferroelectric perovskites," *Phys. Rev. X,* vol. 11, no. 2, 2021.

[27] H. W. Lin, G. Mead and G. A. Blake, "Mapping LiNbO3 phonon-polariton nonlinearities with 2D THz-THz-Raman spectroscopy," *Phys. Rev. Lett.,* vol. 129, no. 20, 2022.

[28] S. Houver, L. Huber, M. Savoini, E. Abreu and S. L. Johnson, "2D THz spectroscopic investigation of ballistic conduction-band electron dynamics in InSb," *Opt. Expr.,* vol. 27, no. 8, 2019.

[29] T. Blank, K. Grishunin, K. Zvezdin, N. Hai, J. Wu, S.-H. Su, J.-C. Huang, A. Zvezdin and A. Kimel, "Two-Dimensional Terahertz Spectroscopy of Nonlinear Phononics in the Topological Insulator MnBi2Te4," *Phys. Rev. Lett.,* vol. 131, no. 2, 2023.

[30] L. Luo, M. Mootz, J. H. Kang, C. Huang, K. Eom, J. W. Lee, C. Vaswani, Y. G. Collantes, E. E. Hellstrom, I. E. Perakis, C. B. Eom and J. Wang, "Quantum coherence tomography of light-controlled superconductivity," *Nat. Phys.,* vol. 19, no. 2, pp. 201-209, 2023.

[31] M. Mootz, L. Luo, J. Wang and I. E. Perakis, "Visualization and quantum control of light-accelerated condensates by terahertz multi-dimensional coherent spectroscopy," *Commun. Phys.,* vol. 5, no. 1, pp. 1-10, 2022.

[32] M.-J. Kim, S. Kovalev, M. Udina, R. Haenel, G. Kim, M. Puviani, G. Cristiani, I. Ilyakov, T. V. A. G. de Oliveira, A. Ponomaryov, J.-C. Deinert, G. Logvenov, B. Keimer, D. Manske, L. Benfatto and S. Kaiser, "Tracing the dynamics of superconducting order via transient third harmonic generation," *arXiV,* 2023.

[33] A. Honold, L. Schultheis, J. Kuhl and C. W. Tu, "Reflected degenerate four-wave mixing on GaAs single quantum wells," *Appl. Phys. Lett,* vol. 52, pp. 2105-2107, 1988.

[34] S. Rajasekaran, E. Casandruc, Y. Laplace, D. Nicoletti, G. D. Gu, S. R. Clark, D. Jaksch and A. Cavalleri, "Parametric amplification of a superconducting plasma wave," *Nat. Phys.,* vol. 12, pp. 1012-1016, 2016.

[35] S. Zhang, Z. Sun, Q. Liu, Z. Wang, Q. Wu, L. Yue, S. Xu, T. Hu, R. Li, X. Zhou, J. Yuan, G. D. Gu, T. Dong and N. Wang, "Revealing the frequency-dependent oscillations in nonlinear terahertz response induced by Josephson current," *Natl. Sci. Rev..*

[36] M. E. Siemens, G. Moody, H. Li, A. D. Bristow and S. T. Cundiff, "Resonance lineshapes in two-dimensional Fourier transform spectroscopy," *Opt. Expr.,* vol. 18, no. 17, pp. 17699-17708, 2010.

[37] A. D. Bristow, T. Zhang, M. E. Siemens, S. T. Cundiff and R. P. Mirin, "Separating homogeneous and inhomogeneous line widths of heavy-and light-hole excitons in weakly disordered semiconductor quantum wells," *J. Phys. Chem. B,* vol. 115, no. 18, pp. 5365-5371, 2011.

[38] A. Liu, S. T. Cundiff, D. B. Almeida and R. Ulbricht, "Spectral broadening and ultrafast dynamics of a nitrogen-vacancy center ensemble in diamond," *Mater. Quantum Technol.,* vol. 1, no. 2, 2021.

[39] A. E. Koshelev and L. N. Bulaevskii, "Fluctuation broadening of the plasma resonance line in the vortex liquid state of layered superconductors," *Phys. Rev. B,* vol. 60, no. 6, 1999.

[40] M. Tinkham, Introduction to superconductivity, Mineola: Dover Publications, 2004.

[41] D. M. Ginsberg, "Calculation of the temperature dependence of the electromagnetic coherence length in superconductors," *Phys. Rev. B,* vol. 7, no. 1, 1973.

[42] G. Deutscher, "The role of the short coherence length in unconventional superconductors," *Condens. Matter,* vol. 5, no. 4, 2020.

[43] B. Liu, H. Bromberger, A. Cartella, M. Först and A. Cavalleri, "Generation of narrowband, high-intensity, carrier-envelope phase-stable pulses tunable between 4 and 18 THz," *Opt. Lett.,* vol. 42, no. 1, pp. 129-131, 2017.

[44] S. Kaiser, "Light-induced superconductivity in high-Tc cuprates," *Physica Scripta,* vol. 92, no. 10, 2017.

[45] B. Liu, M. Forst, M. Fechner, D. Nicoletti, J. Porras, T. Loew, B. Keimer and A. Cavalleri, "Pump frequency resonances for light-induced incipient superconductivity in YBa2Cu3O6.5," *Phys. Rev. X,* vol. 10, no. 1, 2020.

[46] S. Kaiser, C. R. Hunt, D. Nicoletti, W. Hu, I. Gierz, H. Y. Liu, M. Le Tacon, T. Loew, D. Haug, B. Keimer and A. Cavalleri, "Optically induced coherent transport far above Tc in underdoped YBa2Cu3O6+δ," *Phys. Rev. B,* vol. 89, no. 18, 2014.

[47] K. A. Müller and H. Burkhard, "SrTiO3: an intrinsic quantum paraelectric below 4 K," *Phys. Rev. B,* vol. 19, no. 7, 1979.

[48] Y. Wan and N. P. Armitage, "Resolving Continua of Fractional Excitations by Spinon Echo in THz 2D Coherent Spectroscopy," *Phys. Rev. Lett.,* vol. 122, no. 25, 2019.

[49] C. Bao, P. Tao, D. Sun and S. Zhou, "Light-induced emergent phenomena in 2D materials and topological materials," *Nat. Rev. Phys.,* vol. 4, pp. 33-48, 2022.

[50] A. S. Disa, T. F. Nova and A. Cavalleri, "Engineering crystal structures with light," *Nat. Phys.,* vol. 17, no. 10, pp. 1087-1092, 2021.